\documentclass[runningheads]{llncs} 

\usepackage{graphicx}
\usepackage{booktabs} 
\usepackage{graphicx}
\usepackage{subcaption}
\usepackage{soul}
\usepackage{algorithm}
\usepackage{algorithmic}
\usepackage{multirow}
\usepackage{array}
\usepackage{amsmath}
\usepackage[font=small,skip=0.001cm]{caption}
\usepackage{tabularx}
\usepackage{float}
\title{Benchmarking Preprocessing and Integration Methods in Single-Cell Genomics}

\author{Ali Anaissi$^{1,2}$, Seid Miad Zandavi$^{2,3}$, Weidong Huang$^{1}$, Junaid Akram$^{2}$, \\ Basem Suleiman$^{4}$, Ali Braytee$^{1}$ and Jie Hua$^{5}$}

\institute{University of Technology Sydney, Australia \and
University of Sydney, Australia \and 
Broad Institute, United States \and
University of New South Wales,  Australia \and
Shaoyang University, China  \\
\email{ali.anaissi@uts.edu.au, szandavi@broadinstitute.org, weidong.huang@uts.edu.au, Junaid.Akram@uts.edu.au, b.suleiman@unsw.edu.au, ali.braytee@uts.edu.au, steven.hua@mq.edu.au}
}

\begin{document}

\maketitle
\thispagestyle{empty}
\pagestyle{empty}

\begin{abstract}
Single-cell data analysis has the potential to revolutionize personalized medicine by characterizing disease-associated molecular changes at the single-cell level. Advanced single-cell multimodal assays can now simultaneously measure various molecules (e.g., DNA, RNA, Protein) across hundreds of thousands of individual cells, providing a comprehensive molecular readout. A significant analytical challenge is integrating single-cell measurements across different modalities. Various methods have been developed to address this challenge, but there has been no systematic evaluation of these techniques with different preprocessing strategies. This study examines a general pipeline for single-cell data analysis, which includes normalization, data integration, and dimensionality reduction. The performance of different algorithm combinations often depends on the dataset sizes and characteristics. We evaluate six datasets across diverse modalities, tissues, and organisms using three metrics: Silhouette Coefficient Score, Adjusted Rand Index, and Calinski-Harabasz Index. Our experiments involve combinations of seven normalization methods, four dimensional reduction methods, and five integration methods. The results show that Seurat and Harmony excel in data integration, with Harmony being more time-efficient, especially for large datasets. UMAP is the most compatible dimensionality reduction method with the integration techniques, and the choice of normalization method varies depending on the integration method used.

\end{abstract}

\section{Introduction}
\label{s:int}

Technological advances have significantly increased our ability to generate high-throughput single-cell gene expression data\cite{lun2019further}. However, single-cell data often originates from multiple experiments with variations in capturing time, personnel, reagents, equipment, and technology platforms, leading to large variations that can confound biological variations during data integration. scRNA-seq integration\cite{kharchenko2021triumphs,alyassine2024efficient,wu2025simplified} addresses two main issues: generating cell-type feature clusters and determining whether clusters represent actual cell types or result from biological or technological variations, such as specific batch effects or high mitochondrial content. Despite its potential, scRNA-seq integration faces risks, including low-quality cluster identification due to meaningless variations and biased clustering from improper arrangement of similar cell types.

A popular strategy introduced by Haghverdi et al. \cite{haghverdi2018batch} identifies cell mappings between datasets and reconstructs the data in a shared space by finding mutual nearest neighbors (MNNs) \cite{haghverdi2018batch,lun2019further}. This method, while effective in generating a normalized gene expression matrix suitable for downstream analysis, is computationally intensive. To address this, the fastMNN algorithm applies the MNN technique in a PCA-computed subspace, improving performance and accuracy \cite{jolliffe2016principal}. Similarly, Scanorama searches for MNNs in dimensionally reduced regions for batch integration \cite{hie2019efficient}.

scRNA-seq integration analysis typically involves four modules: data normalization, dimensionality reduction, data integration, and result visualization. Numerous algorithms are available for each module, creating a vast number of possible combinations that need evaluation to determine optimal performance. The performance of these combinations depends heavily on dataset size and type, posing a challenge in identifying the best algorithm and parameter settings. This challenge requires significant computational resources, time, and expertise.

This paper addresses this challenge by introducing an empirical evaluation framework to help scientists evaluate scRNA-seq algorithms and choose the best combinations for their datasets. We investigate optimal clustering model combinations for different types of datasets using various evaluation methods. The framework is divided into three parts: data normalization, dimensionality reduction, and data integration. For normalization, we investigate seven core methods: Log Normalization, Counts Per Million (CPM), SCTransform, TF-IDF, Linnorm, Scran, and TTM \cite{lytal2020normalization,zandavi2022disentangling,zandavi2023fotomics}. For dimensionality reduction, we evaluate PCA, UMAP, t-SNE, and PHATE. For data integration, we assess Seurat, Harmony, FastMNN \cite{haghverdi2018batch,lun2019further}, ComBat \cite{johnson2007adjusting}, and Scanorama \cite{hie2019efficient}. We use three evaluation metrics—Silhouette Coefficient Score, Adjusted Rand Index, and Calinski-Harabasz Index—to examine clustering performance and time efficiency.

Our study selects the best models based on evaluation results for each dataset, analyzing reasons for different combinations' performance. We also provide insights into the rules of method selection for different dataset types and sizes, offering data support for future model selection.

The major contributions of our work are as follows:

\begin{enumerate}
    \item We propose an empirical framework systematically assessing various computational strategies for scRNA-seq data integration. This framework includes seven normalization methods, four dimensionality reduction techniques, and five integration methods, providing a holistic approach to scRNA-seq data analysis.
    
    \item Utilizing robust evaluation metrics—Silhouette Coefficient, Adjusted Rand Index, and Calinski-Harabasz Index—we analyze 140 combinations of the methods. This evaluation elucidates performance efficiency and scalability, offering critical insights into their applicability in clustering cell types and aligning datasets from varied sources.
    
    \item Our comparative analysis identifies the most effective combinations of normalization, dimensionality reduction, and integration methods for scRNA-seq data. This provides a strategic roadmap for researchers, facilitating high-fidelity integration of heterogeneous single-cell datasets and enhancing biological insights.
\end{enumerate}

\section{Related Work}
\label{s:related}

Single-cell RNA sequencing (scRNA-seq) has transformed the discovery and characterization of cellular phenotypes, aiding in the identification of biomarkers within the biomedical field \cite{yu2020single}. The foundational principle of scRNA-seq involves measuring gene expression distributions across cell populations, as described by Tang et al. \cite{tang2009mrna}. Since 2014, advancements have significantly reduced sequencing costs and enhanced protocols, broadening its application. scRNA-seq has been pivotal in profiling the molecular regulation of T lymphocytes, leading to new insights into molecular determinants \cite{aldridge2020single}. The Human Cell Atlas (HCA) Global Alliance uses this technology to create a reference map of human tissues, promising advancements in understanding aging, disease, and potential treatments. Future applications extend to cell-based models, cell therapies, and regenerative medicine.

However, scRNA-seq data presents challenges, notably the batch effect, arising from variations in data collection and processing, which can hinder data integration and interpretation. Seurat is widely used for mitigating batch effects and integrating various single-cell data types. It employs Canonical Correlation Analysis (CCA) and anchoring techniques to address gene expression discrepancies through weighted-nearest neighbor analysis \cite{hao2021integrated}. Despite its utility, Seurat’s performance can decline with a high number of batches, particularly when dealing with non-highly variable genes \cite{lakkis2021joint}. To address this, Lakkis et al. introduced CarDEC, a deep learning model enhancing scRNA-seq data by increasing information content while denoising. Peng et al. \cite{peng2021integration} proposed the cFIT method, an unsupervised approach that integrates data from multiple sources with fewer restrictions, improving batch effect correction.

Normalization is crucial for reducing batch effects while preserving biological variation \cite{hafemeister2019normalization}. Techniques like TMM have shown success but can over-correct, prompting recommendations for methods like Linnorm and SCnorm, specifically designed for scRNA-seq \cite{lytal2020normalization}. scRNA-seq data, characterized by high dimensionality, sparsity, and noise, often requires dimensionality reduction to transform it into a lower-dimensional space while preserving meaningful properties. Methods such as PCA, UMAP, t-SNE, and deep count autoencoder (DCA) each have strengths and weaknesses, with UMAP preserving global structures but potentially introducing noise \cite{xiang2021comparison}. Visualization methods like UMAP and t-SNE are intuitive for evaluating integration effectiveness but need quantitative metrics like local inverse Simpson’s index, average silhouette width, and adjusted rand index for rigorous assessment \cite{lazar2013batch}.

\section{Methodology}
\label{Methodology}

We propose a comprehensive framework for the integration of scRNA-seq data, consisting of multiple stages: data preprocessing, dimensionality reduction, data integration, and evaluation of clustering performance. Each stage employs various established methods to ensure robust and accurate results.

Initially, data preprocessing involves normalization using several methods, including Log-Normalization, Counts Per Million (CPM), SCTransform, Term Frequency-Inverse Document Frequency (TF-IDF), Linnorm, Scran, and the Trimmed Mean of M-values (TMM). Following normalization, dimensionality reduction techniques such as Principal Component Analysis (PCA), Uniform Manifold Approximation and Projection (UMAP), t-Distributed Stochastic Neighbor Embedding (t-SNE), and Potential of Heat-diffusion for Affinity-based Transition Embedding (PHATE) are employed to transform high-dimensional data into a lower-dimensional space, facilitating visual inspection and further analysis.

Given the varied performance of dimensionality reduction methods in separating biological clusters and detecting rare cell populations, we systematically assess their effectiveness in conjunction with different scRNA-seq integration methods.

\begin{figure*}[!t]
  \centering
  \includegraphics[width=\textwidth]{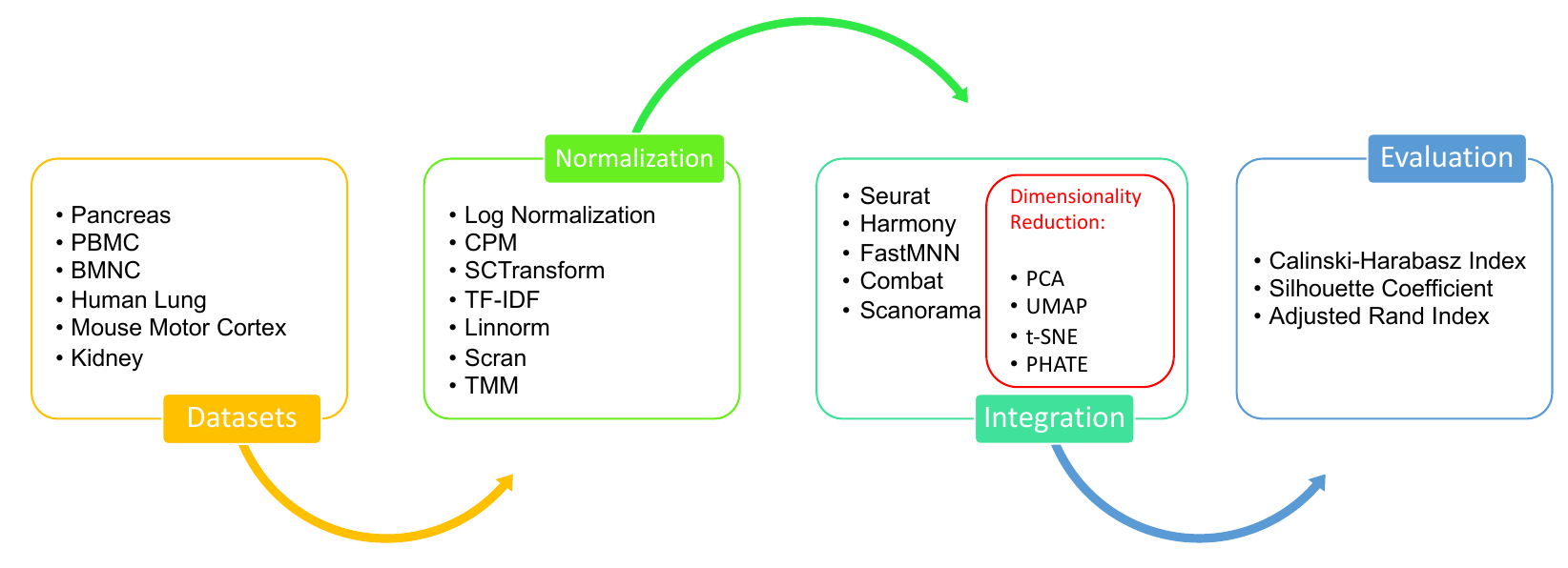}
  \caption{Process Flow of the methods}
  \label{framework}
\end{figure*}

Next, we integrate the processed data and cluster cells using popular methods including Seurat, Harmony, Fast Mutual Nearest Neighbors (FastMNN), Combat, and Scanorama. The results are visualized using DimPlots, and evaluated using the Silhouette Coefficient Score, Calinski-Harabasz Index, and Adjusted Rand Index to measure clustering performance. Figure \ref{framework} illustrates our proposed framework.

\subsection{Normalization Methods}

We have chosen the following normalization methods to investigate as part of our framework:

\begin{itemize}
    \item \textbf{Log Normalization}: This method uses the log function to scale larger values to a smaller interval, improving model accuracy by reducing the impact of large numerical weights.
    
    \item \textbf{Counts Per Million (CPM)}: CPM involves dividing the count columns by their total fragments and scaling by millions, followed by a log transformation. This method is used by Stuart et al. \cite{stuart2019comprehensive} for scaling and filtering scATAC-seq gene matrices before dimensionality reduction.
    
    \item \textbf{SCTransform}: An algorithm for normalization and variance stabilization, SCTransform uses a regularized negative binomial model, constructing a generalized linear model for each gene with sequencing depth as the explanatory variable and UMI counts as the response variable \cite{hafemeister2019normalization}.
    
    \item \textbf{TF-IDF}: A method standard in text analysis, TF-IDF analyzes the importance of genes (words) in cells (documents) by their frequency and inverse document frequency \cite{moussa2018single}.
    
    \item \textbf{Linnorm}: This normalization method uses a linear model and normality to perform accurate statistics and analysis on scRNA-seq datasets, using strictly selected homologous genes as a reference \cite{yip2017linnorm}.
    
    \item \textbf{Scran}: An R package for RNA-seq data analysis, Scran's computeSumFactors method normalizes cell-specific biases by deconvolution \cite{l2016pooling}.
    
    \item \textbf{Trimmed Mean of M-values (TMM)}: TMM uses weighted trimmed mean of log expression ratios to estimate RNA production, normalizing the data by calculating the $M$ and $A$ values, which represent log expression ratios and average expression levels, respectively.
\end{itemize}

\subsection{Dimensionality Reduction Methods}

The following dimensionality reduction methods are investigated as part of our proposed framework:

\begin{itemize}
    \item \textbf{Principal Component Analysis (PCA)}: A linear dimensionality reduction method, PCA transforms correlated variables into a small number of uncorrelated principal components \cite{liu2020visualizing}.
    
    \item \textbf{Uniform Manifold Approximation and Projection (UMAP)}: A non-linear dimensionality reduction technique that preserves more of the global structure of the data compared to other methods, offering excellent runtime performance \cite{schofield2021using}.
    
    \item \textbf{t-SNE}: This method converts high-dimensional data into a lower-dimensional space while maintaining the probability distribution of the data points before and after the reduction. t-SNE uses a t-distribution in the lower-dimensional space to improve separation between clusters \cite{kobak2019art}.
    
    \item \textbf{PHATE}: A visualization method for high-dimensional data, PHATE retains the global structure of the data and shows the information-geometric distance between data points. It is robust to noise and scalable to large datasets \cite{moon2019visualizing}.
\end{itemize}

\subsection{Integration Methods}

Integration methods are essential for removing unwanted technical variation while preserving valid biological variation. We employ the following integration methods for batch correction and data integration:

\begin{itemize}
    \item \textbf{Seurat}: An R package designed for single-cell transcriptome sequencing and analysis, Seurat integrates various types of single-cell data and analyzes heterogeneity from single-cell transcriptomic measurements.
    
    \item \textbf{Harmony}: An efficient algorithm for integrating large single-cell datasets, Harmony starts by clustering cells in a low-dimensional embedding space and iteratively refines these clusters based on a metric that penalizes inappropriate cluster compositions \cite{korsunsky2019fast}.
    
    \item \textbf{FastMNN}: This method corrects batch effects using a modified mutual nearest neighbors (MNN) approach, identifying MNN pairs after dimensionality reduction and correcting batch effects accordingly \cite{zhang2019novel}.
    
    \item \textbf{ComBat}: An empirical Bayesian framework, ComBat corrects batch effects by standardizing data, estimating batch effect parameters, and adjusting data based on these estimates \cite{johnson2007adjusting}.
    
    \item \textbf{Scanorama}: This method integrates single-cell datasets from different technologies using panoramic batch correction and integration. It employs SVD for dimensionality reduction and constructs a nearest neighbor graph for integration \cite{hie2019efficient}.
\end{itemize}





    


    

\subsection{Data Analysis}

The Wilcoxon Rank-Sum Test is employed for data analysis. This non-parametric test compares the distribution of two independent samples to determine if they come from the same distribution. After calculating the Silhouette Coefficient, Calinski-Harabasz Index, and Adjusted Rand Index scores for each dataset, we rank the methods and apply the Wilcoxon Rank-Sum Test to identify the best normalization, dimensionality reduction, and integration approaches \cite{lamorte2017mann}.

By following this comprehensive methodology, we aim to systematically assess the performance of various normalization, dimensionality reduction, and integration methods in scRNA-seq data analysis, ensuring robust and accurate results across different datasets.

\begin{figure}[!t]
  \centering
  \begin{subfigure}[b]{.48\textwidth}
    \centering
    \includegraphics[width=\textwidth]{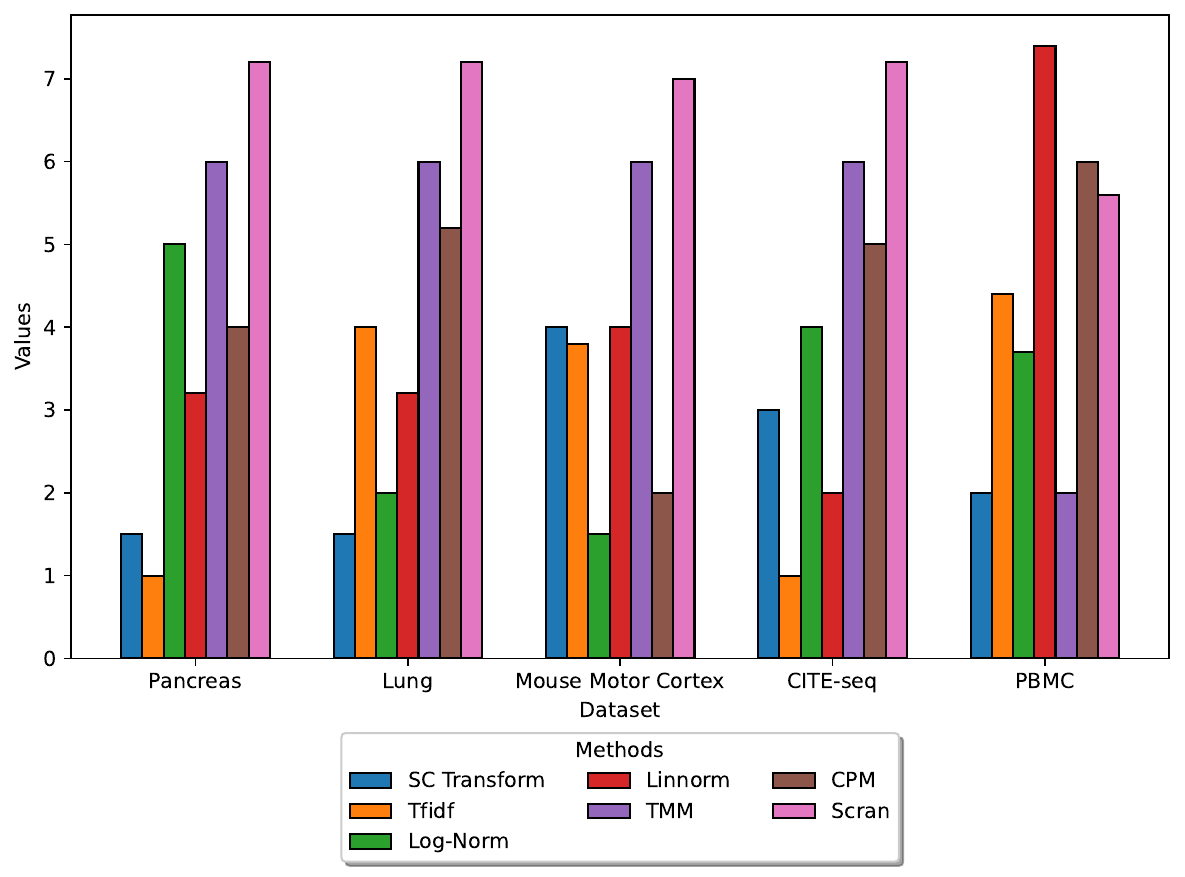}
    \caption{Wilcoxon Rank-Sum Test of Normalization Methods.}
    \label{fig_rank_norm}
  \end{subfigure}
  \hfill
  \begin{subfigure}[b]{.48\textwidth}
    \centering
    \includegraphics[width=\textwidth]{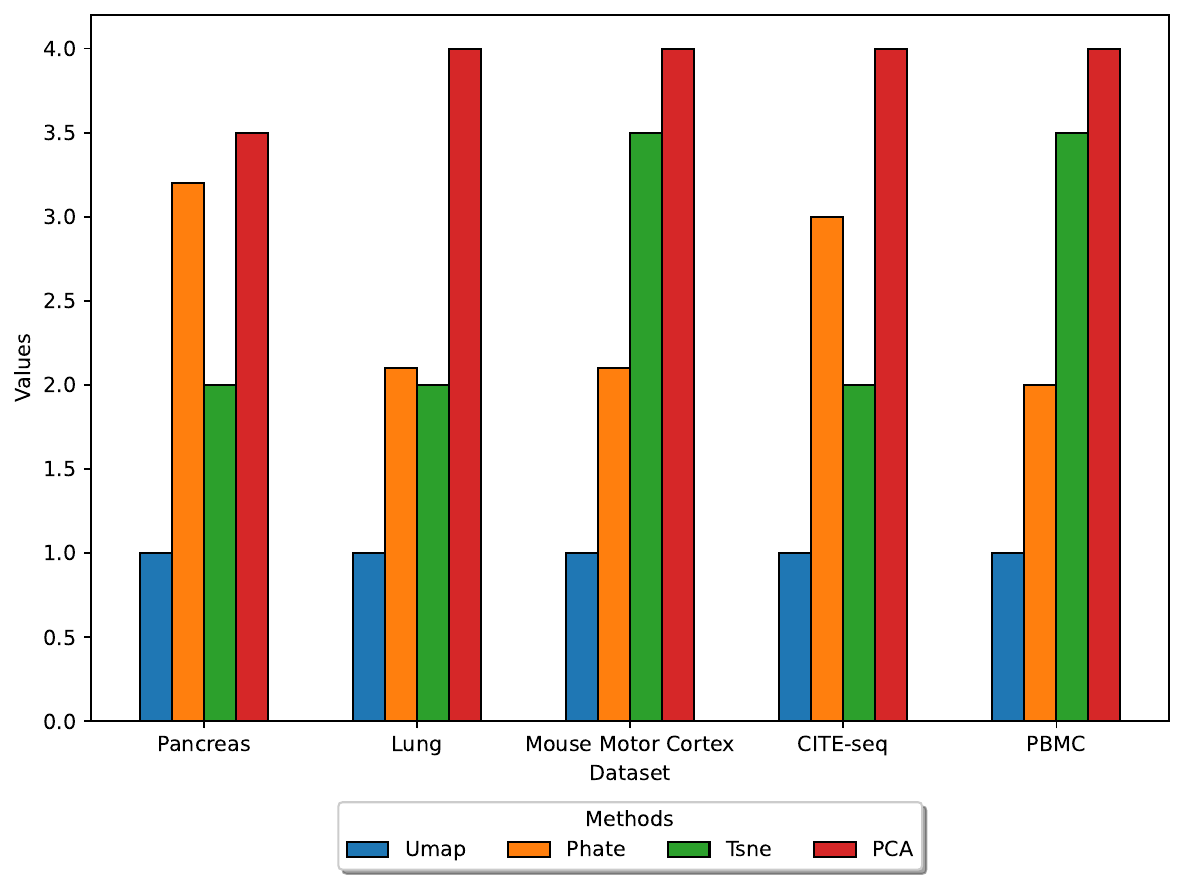}
    \caption{Wilcoxon Rank-Sum Test of Dimension Reduction Methods.}
    \label{fig_rank_dim}
  \end{subfigure}
  \hfill
  \begin{subfigure}[b]{.48\textwidth}
    \centering
    \includegraphics[width=\textwidth]{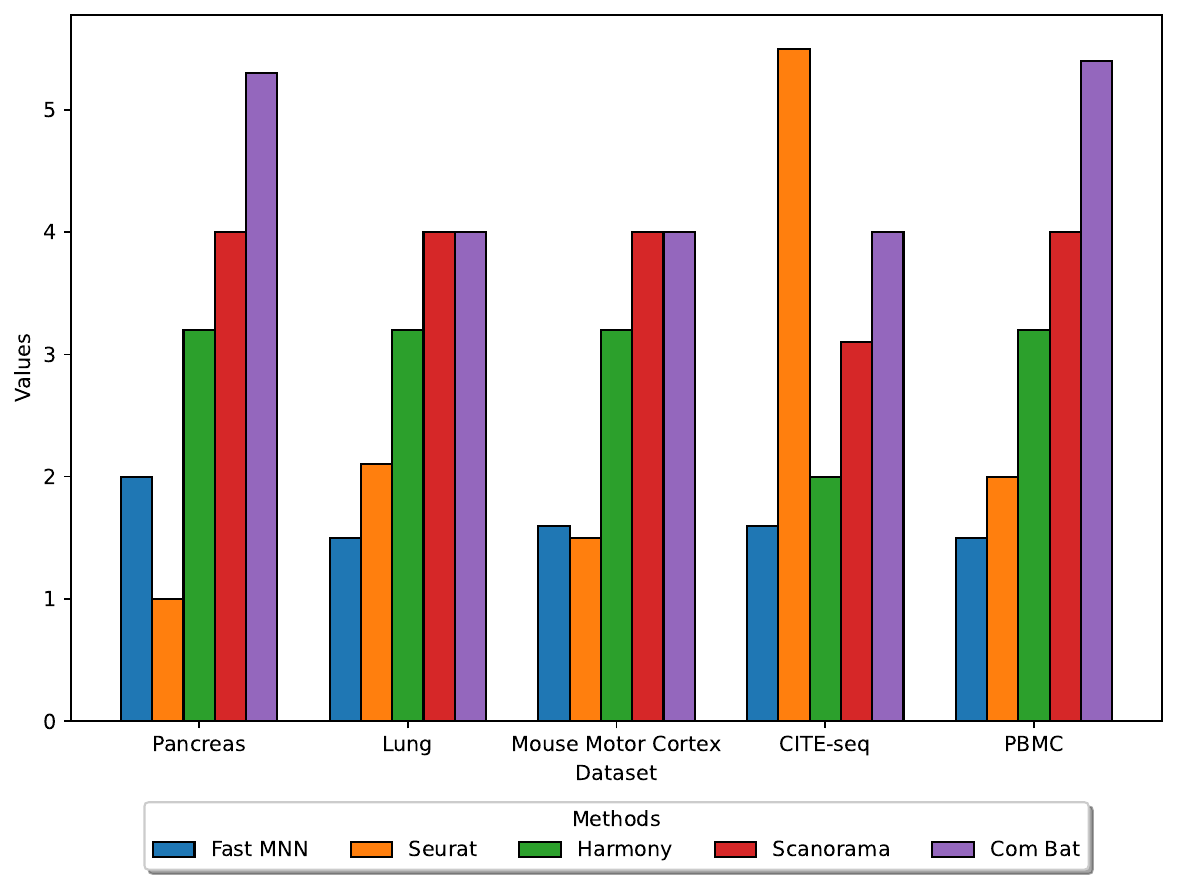}
    \caption{Wilcoxon Rank-Sum Test of Integration Methods.}
    \label{fig_rank_integration}
  \end{subfigure}
  \caption{Wilcoxon Rank-Sum Tests for Various Methods}
  \label{fig_combined}
\end{figure}

\begin{table*}[!t]
\centering
\tiny
\caption{Summary of Evaluation Results}
\label{table:summary_evaluation_results}
\begin{tabular}{|c|c|c|c|c|c|c|c|c|}
\hline
 \textbf{Method} & \textbf{CH Sum} & \textbf{Silh. Sum} & \textbf{ARI Sum} & \textbf{CH Rank} & \textbf{Silh. Rank} & \textbf{ARI Rank} & \textbf{Rank Sum} & \textbf{Final Rank} \\

\hline \hline
\multicolumn{9}{|c|}{\textbf{Summary Normalisation Results}} \\
\hline
Log-Norm & 995503.848 & 32.3981 & 63.158689 & 5 & 7 & 7 & 19 & 1 \\
CPM & 904642.351 & 28.8685 & 58.4329301 & 3 & 3 & 3 & 9 & 5 \\
SCTransform & 1130895.78 & 31.8076 & 60.8940582 & 7 & 6 & 6 & 19 & 1 \\
TFIDF & 1004035.56 & 31.4413 & 60.7435297 & 6 & 5 & 5 & 16 & 3 \\
Linnorm & 935391.89 & 29.7849 & 59.721815 & 4 & 4 & 4 & 12 & 4 \\
SCRAN & 743686.811 & 9.28917 & 40.5276838 & 1 & 1 & 1 & 3 & 7 \\
TMM & 813891.279 & 27.62767 & 58.2278707 & 2 & 2 & 2 & 6 & 6 \\
\hline \hline
\multicolumn{9}{|c|}{\textbf{Summary Dimension Reduction Results}} \\
\hline
PCA & 203086.58 & 26.3973 & 83.709358 & 1 & 1 & 1 & 3 & 4 \\
UMAP & 2694239.70 & 67.8274 & 112.5972185 & 4 & 4 & 4 & 12 & 1 \\
TSNE & 1205768.39 & 56.3184 & 101.296 & 2 & 3 & 3 & 8 & 2 \\
PHATE & 2380788.68 & 39.88214 & 94.727 & 3 & 2 & 2 & 7 & 3 \\
\hline \hline
\multicolumn{9}{|c|}{\textbf{Summary Integration Results}} \\
\hline
Seurat & 1144219.94 & 47.57594 & 85.112 & 2 & 4 & 4 & 10 & 2 \\
Harmony & 1525235.92 & 39.9727 & 82.208 & 4 & 3 & 3 & 10 & 2 \\
FastMNN & 1538792.95 & 50.8561 & 96.145 & 5 & 5 & 5 & 15 & 1 \\
Combat & 846507.87 & 21.849 & 59.1045765 & 1 & 1 & 1 & 3 & 5 \\
Scanorama & 1251230.84 & 30.9551 & 79.137 & 3 & 2 & 2 & 7 & 4 \\
\hline
\end{tabular}%

\end{table*}

\section{Experiments and Results}
\label{ResuDiscuss}

This section discusses the  model performance evaluation along with the time efficiency evaluation on five different datasets. 

\subsection{Datasets}\label{DataCollection}
We conducted experiments on five RNA gene sequences datasets which are described as follows:

\begin{itemize}

\item {\textbf{Pancreas}} dataset is a combination of different pancreas scRNA-seq datasets from eight studies using  five different techniques.  It is integrated into a single cell  using Seurat\cite{stuart2019comprehensive}.
\item {\textbf{Peripheral blood mononuclear cell (PBMC)}} dataset is generated based on the eight volunteers enrolled in an HIV vaccine trial. It takes three-time point samples at days 0, 3, and 7 following vaccination to form 24 samples, which is processed by using the CITE-seq technique to produce RNA and ADT.
\item {\textbf{CITE-seq}} dataset contains 30,672 samples of human bone marrow mononuclear cells (BMNC) and 25 antibodies, which were derived from eight individual donors. BMNC dataset, generated by the Human Cell Atlas, gains two assays, RNA and antibody-derived tags (ADT).
\item {\textbf{Human Lung cells}} dataset\cite{vieira2019cellular} contains 58 molecular cell types from 65,662 human lung and blood cells, including bronchi, bronchiole, alveoli and circulating blood.
\item {\textbf{Mouse motor cortex}} dataset is referenced from Yao et al\cite{yao2021transcriptomic} which analyzes adult mouse isocortex and hippocampal formation to gain transcriptomic and epigenomic atlas from 12 individual mice.

\end{itemize}

\subsection{Method Performance Evaluation}

As a result of the evaluation performance of each dataset, a variety of different method combinations were determined to be the most effective. To combine the rankings across all metrics, we used the Wilcoxon Rank-Sum approach to rank methods based on each of the CH, SC, and ARI metrics. A lower rank-sum score indicates better performance when it comes to calculating the height of ridge-lines across different datasets. Methods are ranked from top to bottom based on the sum of their rank scores for the six data sets, with the top-performing methods appearing at the top. Additionally, the datasets on the x-axis are sorted in ascending order of the size of the dataset, which is calculated by features across samples.

In terms of the normalization method (Fig~\ref{fig_rank_norm}), SCTransform, Log Normalization and TF-IDF come out as the top three methods with the most remarkable overall performance. These methods were ranked among the top three in four datasets, including Pancreas, PBMC, CITE-seq, and Mouse Motor Cortex. SCTransform produced the highest quality normalization results for Pancreas and CITE-seq, while poor results were obtained for Mouse Motor Cortex. Log Normalization ranked within the top four in all datasets except for Lung. Also, TF-IDF scored highest in Mouse Motor Cortex and lung, and best three in PBMC. TF-IDF and SCTransform performed well when dealing with small datasets, while SCTransform was also able to run the larger dataset successfully. Fig~\ref{fig_rank_norm} shows that log normalization performed better for large datasets as a decreasing tendency.

Typically, batch integration is evaluated visually by examining t-SNE or UMAP plots, whilst our experiments also use PHATE plots. Fig~\ref{fig_rank_dim} depicts UMAP's significant superiority over other methods of dimension reduction. UMAP consistently ranks first across all data sets without limiting the size of the data set, which proves that UMAP has a beneficial effect on the dimensionality reduction process of scRNA-seq integration. PHATE comes in second place in the overall results of evaluation metrics, and its evaluation performance is significant across most datasets. TSNE and PCA are the most under-performing methods for dimensionality reduction notably PCA is the least effective across all five datasets.

The assessment metrics for evaluating the integration methods relating to the different datasets are provided in Section \ref{ResuDiscuss} which outlines in detail the ranking of each method. As shown in Fig~\ref{fig_rank_integration}, the computed rank sum ranked FastMNN as the top method, with Seurat and Harmony ranked second (See table~\ref{table:summary_evaluation_results}). FastMNN produces the best results on mouse motor, pancreas, and PBMC datasets, but it does poorly on CITE-seq. Scanorama method was the least effective compared to other methods.
It can be concluded that the FastMNN method is suitable for handling datasets of any size. Generally, Seurat performs better with smaller datasets, whereas Harmony performs well with large and small datasets.

\subsection{Time Efficiency Evaluation}

\begin{figure}[!t]
  \centering
  \begin{subfigure}[b]{.48\textwidth}
    \centering
    \includegraphics[width=\textwidth]{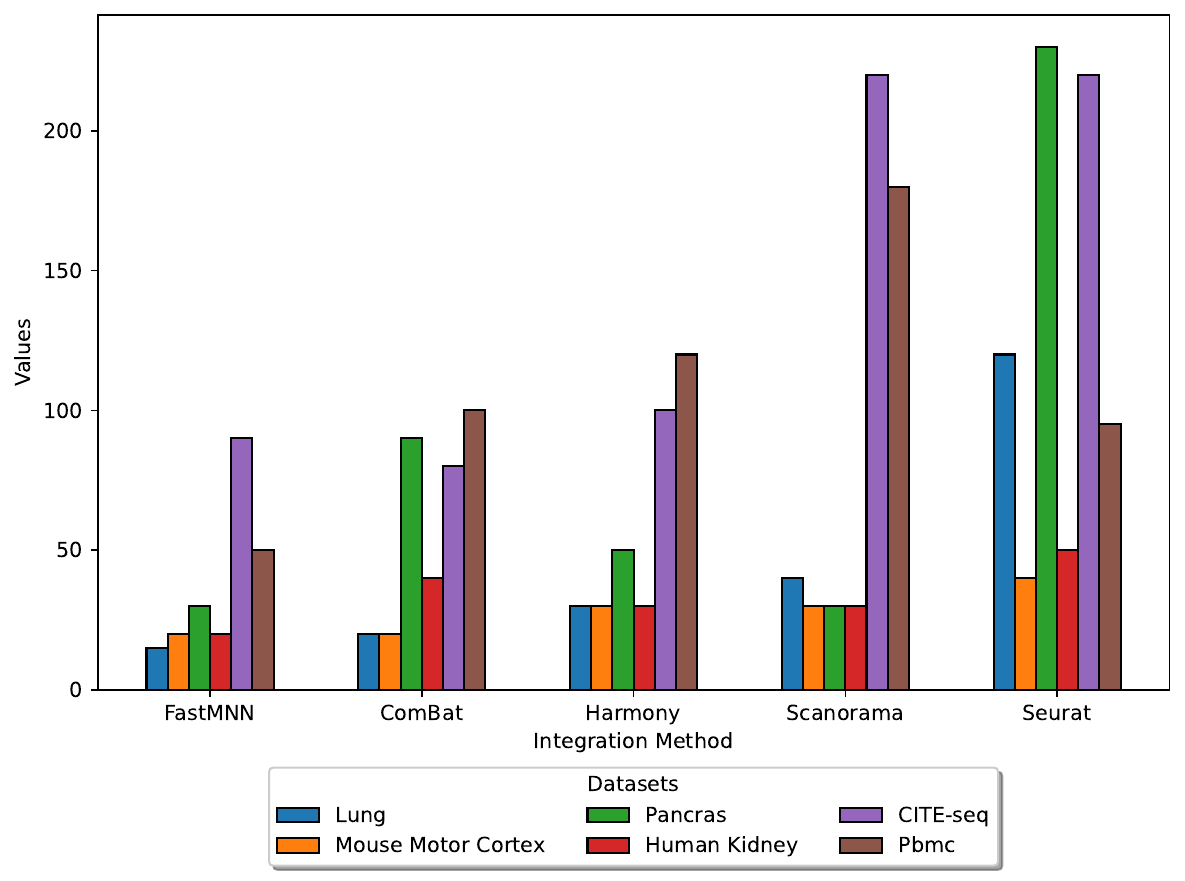}
    \caption{Comparison of Integration Methods Across Datasets.}
  \end{subfigure}
  \hfill
  \begin{subfigure}[b]{.48\textwidth}
    \centering
    \includegraphics[width=\textwidth]{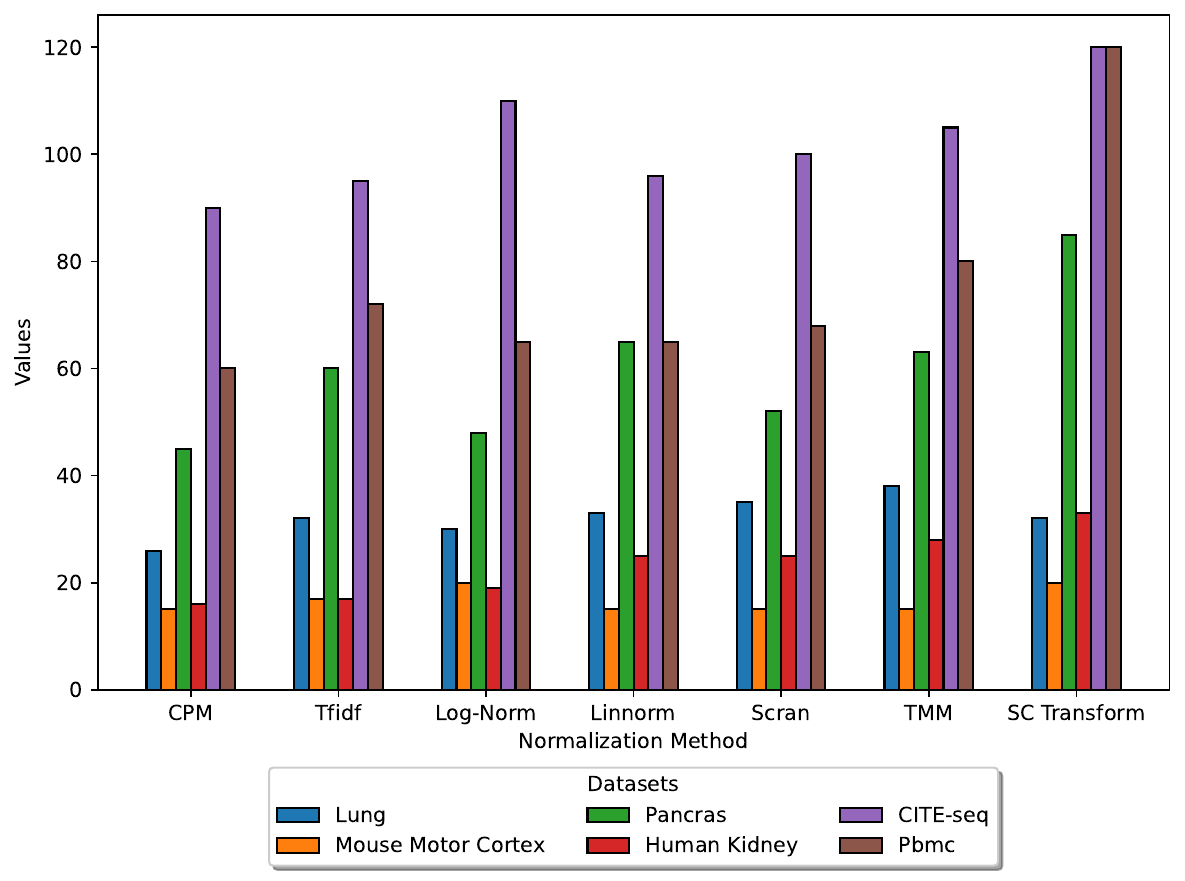}
    \caption{Comparison of Normalization Methods Across Datasets.}
  \end{subfigure}
  \caption{Rank of Running Efficiency}
  \label{fig_discussion}
\end{figure}

Computational time is another important factors to evaluate the pros and cons of a model. Fig~\ref{fig_discussion} shows the comparative analysis  of integration algorithms and standardization methods in terms of running efficiency. Because the file size of different datasets, operating environment and hardware equipment conditions have significant differences. To avoid interference, only the differences of methods between the same datasets are compared. For the data integration method, although the performance of Seurat is the best for clustering, the Seurat method takes a long time. Overall, the least time-consuming method is FastMNN, and the clustering performance is also relatively good, which means this method is more ideal.

For the data normalization method, there is no big difference between the different methods, but the SCTransform method takes a long time. However, the longer time can be accepted because of its excellent performance. Besides, Linnorm and SCTransform have almost the same good performance in the analysis of the model clustering performance, however in terms of efficiency, Linnorm has a more tremendous advantage, so Linnorm is better as a standardized method. In a nutshell, the optimal model for different data sets needs to be comprehensively determined. The above discussion can only be used as a reference.

 
\section{Conclusion}\label{Conclusion}
 
We present a comparative analysis to evaluate the performance of workflows composed of different pre-processing methods and integration methods on six datasets. It can be seen from the result that it is necessary to choose different workflows according to the size and other characteristics of different datasets. In addition, using the subset of it for large datasets can greatly improve the efficiency of comparing different integration methods. We conduct experiments based combinations of seven normalization methods, four dimensional reduction methods, and five integration methods. Our results demonstrated that for the data integration module, the clustering performance of Seurat and Harmony are more prominent, but the time efficiency of Harmony was better. At the same time, the performance of Seurat for small data sets is superior. For the dimensionality reduction module, the UMAP method shows promising results in compatibility with the integration methods. Due to its significantly shorter computational time, FastMNN is recommended as the first method to try, with the other methods as viable alternatives.

%
%
%
 \bibliographystyle{splncs04}
 \bibliography{ref}

\end{document}